\documentclass{ws-ijmpa}
\usepackage[super,compress]{cite}
\usepackage{graphicx}
\usepackage{hyperref}

\newcommand{\onu}{${0\nu\beta \beta}$}
\newcommand{\tnu}{${2\nu\beta \beta}$}

\newcommand{\QBB}{${Q_{\beta\beta}}$}
\newcommand{\al}{$\alpha$}
\newcommand{\be}{$\beta$}
\newcommand{\ga}{$\gamma$}
 
\newcommand{\TL}{$^{208}\mathrm{Tl}$}

\newcommand{\cpy}{counts keV$^{-1}$kg$^{-1}$ y$^{-1}$}
\newcommand{\cpyI}{counts kg$^{-1}$ y$^{-1}$}

\begin{document}
\markboth{Fabio Bellini}{Potentialities of the future technical improvements in the search
of rare nuclear decays by bolometers}

%
\catchline{}{}{}{}{}
%

\title{Potentialities of the future technical improvements in the search
of rare nuclear decays by bolometers}

\author{Fabio Bellini}

\address{Sapienza Universit\`a di Roma, P.le Aldo Moro 2
Roma, 00185,Italy}
\address{INFN, Sezione di Roma, P.le Aldo Moro 2, 00185, Rome, Italy\\
fabio.bellini@roma1.infn.it}

\maketitle

\begin{history}
\received{Day Month Year}
\revised{Day Month Year}
\end{history}

\begin{abstract}
Bolometers are cryogenic calorimeters which feature excellent energy resolution, low energy threshold, high detection efficiency,  flexibility in choice of materials, particle identification capability if operated as hybrid devices.
After thirty years of rapid progresses, they represent nowadays a leading technology in several  fields: particle and nuclear physics, X-ray astrophysics, cosmology. However, further and substantial developments are required  to increase the sensitivity to the levels envisioned by future researches.  A  review  of the  challenges to be addressed and potentialities of bolometers  in the search for rare nuclear decays is given, with particular emphasis to the neutrinoless double beta decay physics case.

\end{abstract}

\ccode{PACS numbers: 07.20.Mc, 23.40.-s, 23.60.+e,14.60.Pq}

\tableofcontents

\section{Introduction}	
\label{sec:intro}
The need for improved energy resolution and  sensitivity to smaller energy depositions led to the proposal to use  cryogenic techniques as an instrument for detecting low energy particles.\\
In 1984 three different groups suggested low temperature detectors  as an instrument  to investigate fundamental problems in nuclear and astroparticle physics.
 Fiorini and Niinikowsky \cite{Fiorini:1983yj} proposed cryogenic calorimeters for the study of the double beta decay and  of the neutrino mass measurement. Drukier and Stodoloky \cite{drukier1} studied the use of superconducting detectors for the search for the coherent neutrino scattering off nuclei. Finally McCammon and coworkers \cite{McCammon} indicated X-ray astrophysics as a possible field where cryogenic devices  might have an important impact.\\
In the last 30 years, Low Temperature Detectors (LTD) experienced a very rapid growth and reached a level of  maturity and versatility such to represent  one of the leading technology in different research fields nowadays.
Examples are: dark matter \cite{pirroreview}, neutrinoless double beta decays  \cite{Poda:2017jnl}, neutrino mass measurement \cite{Nucciotti:2015rsl},  rare nuclear decays and processes \cite{demarcillac}, X-ray astrophysics \cite{Ullom}, cosmological microwave background precision measurements \cite{pirroreview}.\\
Nevertheless to face the challenges imposed by the aforementioned researches and increase the sensitivity of the current experiments, further and substantial technological developments are necessary.\\ 
This review describes the potentialities of the future technical improvements in the search for rare nuclear decays by bolometers. 
Although the term bolometer was originally used for a detector measuring the intensity of electromagnetic radiation by heating up the detector itself, through the text it will be used as synonymous of  cryogenic calorimeter.

This paper  is organized as follows: in section  \ref{sec:herc} the basic principles of cryogenic detectors are summarized together with the main advantages and drawbacks compared to traditional devices. In section \ref{DBD}  the double beta decay physics case is analyzed: actual limits and future developments are discussed. In particular section \ref{BAPI} describes the effort for the implementation of active background identification techniques while needs for reduction of the environmental  radioactivity is discussed  in section \ref{err}. Finally section \ref{rare}  highlights the use of bolometers for other rare \al\ and \be\ nuclear decays as well as electron capture processes.
\section{High Energy Resolution Calorimeters}	
\label{sec:herc}

\subsection{Conventional Calorimeters}	
\label{sec:conv}
A calorimeter is a device which is sensitive to the energy deposited by a single particle. \\
All conventional calorimeters share the same principle: a ionizing particle interacting with a solid medium, deposits part of its energy into the medium itself. The released energy $E$ produces out-of-equilibrium excitation quanta: electron-hole pairs or photons. The quanta are collected as completely as possible before they decay into an undetectable channels. 
To get good energy resolution,  the detector response must be uniform throughout the detection volume so that the fraction of energy released into the desired channel and the  collection efficiency are the same for all events.
The number of excited quanta is proportional to $E$ and inversely proportional to the mean energy necessary to produce each  of them $w$; 
poissonian statistical fluctuations in the  number of created quanta represent the ultimate limit of the technology.\\
When energy resolution matters, the choice in conventional detectors is limited to germanium or silicon devices. 
For example, in silicon, $w$=3.6  eV  and the best energy resolution obtained in semiconductor X-rays detectors is about 125 eV Full With Half Maximum (FWHM) at 6 keV \cite{quaglia}.
The excitation energy   $w$ is about three times the band gap $E_g$. The presence of  several modes with an excitation energy less than $E_g$  and momentum conservation, that requires lattice vibration excitations (phonons),  implies that 70\% of the energy goes into  undetectable channel.
The statistical contribution to the energy resolution $\Delta E_{rms}$ is:
\begin{equation}
\Delta E_{rms} = \sqrt{wFE},
\label{risoSi}
\end{equation}
where F is the Fano factor \cite{Fano,Klein}, which reduces the overall random spread when multiples excitation mechanisms play a role.
On the other hand, the maximum phonon energy in Si is only 60 meV; much more phonons than electron-hole pairs are produced. The possibility to detect such phonons overcomes the limit imposed on the energy resolution by poissonian fluctuations and allows much smaller energy depositions to be detected.\\  
This is the rationale behind the cryogenic calorimeter technique.\\

\subsection{Bolometers}	
\label{sec:basic}
A bolometer is solid state device composed by an absorber, connected through a thermal link to a heat sink,  and equipped with a temperature sensor (thermometer) for the conversion of phonons into an electrical signal.
 Different thermometers are commonly used depending on the specific application:  high resistivity  doped semiconductors (Neutron Transmutation Doped Thermistors \cite{Hallerf, Haller}), paramagnetic sensors (Metallic Magnetic Calorimeters \cite{porst}), superconducting materials sensors (Kinetic Inductance Detector \cite{day}, Transition Edge Sensors \cite{Ullom}, Superconducting Tunnel Junctions \cite{Enns}, Superheated Superconducting Granules \cite{Enns}).
Several reviews on bolometers have been published  \cite{Enns,bolo-review1,bolo-review2,Twerenbold,McCammon200411,Ullom} which include a very general description of different LTDs making using of  different sensor approaches. Here we recall only the basic principles.

Bolometers can be operated as equilibrium or  non-equilibrium devices. In the former  all the released energy $E$ degrades into heat while  in the latter out-of-equilibrium (ballistic) phonons are collected. 

In the most simple model, $E$ is fully thermalized and  the  temperature variation is $\Delta T=E/C(T)$ where $C(T)$ is the detector  heat capacity at a working temperature $T$.
In order to have the highest  temperature variation, $C$ must be as low as possibile. This leads to the necessity  to operate the detector at temperatures well below one Kelvin and to select materials in order to avoid contributions that increase the heat capacity. 
Several material-related characteristics contribute to the specific heat: the lattice contribution that is proportional to $({T/T_D})^3$ where $T_D$ is the Debye temperature of the material; the electron contribution that depends on ${T/T}_F$ where $T_F$ indicates the Fermi temperature; the paramagnetic component which is proportional to $1/T$. It is clear that the paramagnetic contribution is very dangerous, but also the use of conductors could be limited by the specific heat of electrons.
For a superconductor the electron heat decreases as  $\exp(-2T_c/T)$, where $T_c$ is the superconducting critical temperature. For specific applications, superconductors could represent absorbers of interest in which the excited quanta  are the quasi-particles induced by the breaking of Cooper pairs. 

Typical phonons excitations are limited by the  Debye cut frequency and lie in the range of tens of meV for most materials. The ultimate energy resolution could be therefore very high and limited  only by thermodynamic fluctuations due to the random exchange of phonons with the thermal bath. It has been shown in Ref. \citen{bolo-review2} that  $\Delta E_{rms}$ is: 
\begin{equation}
\Delta E_{rms} =  \sqrt{\xi C_0K_BT_0^2}, 
\label{risoBo}
\end{equation}
where $K_B$ is the Boltzmann constant, $T_0$ is the heat sink temperature and $\xi$ is a parameter that depends on the thermometer characteristic which is  one  in the ideal case but can reach values up to ten.
Using  Eq. \ref{risoBo} a resolution of the order few eV can be reached. In reality  several  contributions can deteriorate  the resolution: Johnson noise of the sensor and polarization network, phonon noise due to the temperature gradient, electronic noise of the amplifier, microphonic noise, metastable electron-holes state or long lived non-thermal phonons. Anyhow, using a suitable thermometer and with an appropriate electronic readout, energy resolutions of few eV are reachable.  \\
It's remarkable that $\Delta E_{rms}$ does not depend on $E$, on the thermal conductance G of the thermal link and of the detector time constant $\tau$=G/C. The feature remains 
valid even including signal and noise power spectra and more refined analyses if the assumption of fully thermalization is achieved \cite{bolo-review2}. This opens the window to operate very massive detectors with different absorber materials provided they are kept at sufficiently low temperature.\\
On the other hand, cryogenic particle detectors sensitive to ballistic phonons, are faster relative to equilibrium devices, since thermal equilibrium often takes a very long time to establish at low temperatures, and with no restrictions on the equilibration time, they offer even more flexibility in choice of materials.  They are  subject to branching statistics as well as ionization detectors but the number of excited quanta is much larger.
They may suffer  from position dependence and/or the lifetime and detection efficiency of  excitations but for applications that require large volumes of dielectric material and do not need exceptionally good energy resolution, the speed advantage may outweigh other considerations.

\subsubsection{Strenghts and Weaknesses}	
\label{sec:limitation}
To summarize, the main advantages of bolometers compared to the well established semiconductor ionization technology are:
\begin{itemlist}
\item better energy resolutions;
\item enhanced sensitivity to low energy release;
\item wide flexibility  of materials usable for the absorber. This  characteristic is of primary importance  when a particular isotope is needed as the absorber or the source of particles.
\end{itemlist}

\noindent Despite these advantages, bolometers show limitations and practical challenges. 

\begin{itemlist}
\item They need very a complicated apparatus necessary to maintain very low temperatures. Despite the improvements in cryogenic techniques, the size of the experimental volume is limited to a cubic meter. 
Moreover the cryostat must be very stable since the signal is represented by a very tiny temperature rise  (hundreds of $\mathrm{\mu}$K for one MeV of released energy $E$) compared to the thermic bath and temperature fluctuations may limit the detector response.
Furthermore unwanted noise, introduced by the cryogenic apparatus itself, could jeopardize the excellent energy resolution and affect the energy threshold.

\item The slowness of the bolometric response, that can extends up to several seconds for equilibrium devices, represents a  concern in the search for rare processes. Pile-up events, induced by cosmic rays and by natural radioactivity in the bolometer itself and surrounding material, requires operation in deep underground sites, protection against external radioactivity and a careful selection of radio-pure materials for the detector itself.

\item  In rare nuclear processes  a small signal must be resolved over a large background. 
 Full thermalized  bolometers are almost equally sensitive to any kind of particle, despite the way energy is released. Electrons, \al\ particles, and nuclear recoils, depositing the same amount of energy in the detector, produce a pulse with the same amplitude and shape.\\ In addiction  the response  is unaffected by the impact point of the event. If this feature allows for an excellent energy resolution, it makes however impossible to distinguish bulk from near surface particle interactions. In other words  bolometers don't have a dead layer at the surface, they are fully sensitive in their volume.\\ The lack of particle identification and  the impossibility to tag surface events makes the external background reduction a paramount concern. This represents the most serious limitation that searches, carried out with bolometric techniques, are facing.

\end{itemlist}

\subsubsection{Hybryd bolometers} 	
\label{sec:hybryd}
To overcome the aforementioned limits, hybrid bolometers were developed, in which a double readout is exploited. The temperature increase is measured in parallel with ionization, scintillation or Cherenkov light detection. This possibility permits the discrimination between events that release energy with different efficiencies in different detectable channels. 
Typical examples are neutrons which interact mainly by nuclear recoils, with only a very small fraction of energy that  goes into the ionization channels or \al\ particles that are quenched in the scintillation channel.
This idea was initially developed for dark matter searches  and lead to realization of very sensitive detectors for both heat-ionization and heat-scintillation devices \cite{cdms,Agnese:2014aze,Agnese:2013ixa,Armengaud:2016cvl,Angloher:2015ewa,Angloher:2014myn,Angloher:2016ooq,Angloher:2016hbv}.

Recently hybrid bolometers come to play an important role in the search for Majorana neutrinos through the neutrinoless double beta decay. The discovery potential  of future experiment is closely  related to successful implementation of this technology.

\section{Double Beta Decay}	
\label{DBD}
\subsection{The \onu\ physics case}
The neutrinoless double beta decay \onu\  \cite{Furry} is a transition, in which a nucleus (A,Z) decays into its isobar (A,Z+2) with the simultaneous emission of two electrons. Both the parent and the daughter nucleus must be more bound than the intermediate one (A,Z+1) in order to  avoid  the occurrence of the sequence of two single beta decays. Such a condition, due to the pairing term,  is fulfilled in nature for 35 even-even nuclei \cite{giuntibook}. \\
This process violates the lepton number by two units; it's not allowed by the Standard Model of interactions but it's envisaged in many of its extensions in which neutrinos are their own antiparticles \cite{giuntibook}. Its discovery would ascertain unambiguously the nature of neutrinos as Majorana fermions \cite{PhysRevD.25.2951}, would constrain the absolute neutrino mass scale and provide support to leptogenesis theories \cite{Luty:1992un}.
In the standard paradigma \cite{giuntibook,pdg} the decay is mediated only by the exchange of three virtual light neutrinos between two charged weak interaction vertices. The chirality mismatch imposed by the V-A structure of the ElectroWeak theory leads to an amplitude proportional to a linear combination of the three neutrino masses. The absolute value of the neutrino masses is unknown yet but their sum is constrained to be less than 0.66 eV at 95\% C.L.  from cosmological observations \cite{pdg,Cremonesi:2013vla}. On the other hand, the squared mass differences are well measured  from neutrino oscillation experiments.\\
Three possible orderings are therefore conceivable: normal hierarchy (NH), in which  m$_{\nu1}$ \textless m$_{\nu2}$ \textless m$_{\nu3}$, inverted hierarchy (IH) where m$_{\nu3}$ \textless m$_{\nu1}$ \textless m$_{\nu2}$, and the quasi-degenerate hierarchy (QD), for which  mass differences are tiny compared  to their absolute values. \\
Being a second-order weak interaction process and due the smallness of neutrinos masses, extraordinary long lifetimes ($\tau$\textgreater10$^{25}$yr) are expected for the \onu\ decay.\\
Despite decades of experimental search it has not been observed so far. 
Actual limits on the mean lifetimes $\tau$ are in the range of $10^{24-26}$ yr \cite{Cremonesi:2013vla,DellOro:2016tmg}; running experiments are deeply probing the QD parameter space and  have the possibility to start to scan the IH region \cite{Cremonesi:2013vla}. \\
The goal of the next generation experiments is to completely  cover the IH mass scheme and to have a high chance to assess the neutrino nature in their first operational stages \cite{Agostini:2017jim}.\\
The main signature of the \onu\ decay is a peak in the sum energy spectrum of the electrons at the transition energy of the reaction (commonly referred as \QBB). 
A typical \QBB\  for nuclei of experimental interest lies in the two-three MeV energy range. 
The signal peak must be resolved  on top of  a continuum background induced by natural and anthropogenic radioactive decay chains and cosmogenic induced activity.
Consequently, the main task in \onu\ searches is to decrease the background in the Region Of Interest (ROI).\\
The requirements to achieve the IH coverage ($\tau \sim 10^{27-28}$ yr) descends consequently: about a few thousand moles of the isotope under study (several hundreds of kgs) must be measured in combination  with a background close to zero at the ton $\times$ year exposure scale and a FWHM energy resolution better than $0.5$\% \cite{Cremonesi:2013vla,Artusa:2014wnl}.\\
Bolometers are natural candidate detector for this purpose.
They show excellent energy resolution,  high detection efficiency thanks to the source-equal-detector approach, scalability to the ton scale and  can be made of  different materials allowing the search in several candidate nuclei.\\
The state of the art in the bolometric search for the \onu\ is represented by the CUORE (Cryogenic Underground Observatory for Rare Events) experiment \cite{Artusa:2014lgv}.
CUORE recently demonstrated  \cite{Alduino:2017ehq}  that  a thousand TeO$_2$ bolometers can be successfully operated at a temperature of about ten mK studying the decay of the $^{130}$Te ($Q_{\beta\beta}$ $\sim$ 2527 keV \cite{Redshaw:2009zz, scielzo09,Rahaman:2011zz}).
Despite the improvement compared to its predecessor CUORICINO  \cite{Andreotti:2010vj} and CUORE-0 \cite{Aguirre:2014lua, Alfonso:2015wka,Alduino:2016vjd}, the CUORE background is  currently of the order of $10^{-2}$ \cpy \cite{Alduino:2017ehq} as expected from simulations \cite{Alduino:2017qet}.\\
Its sensitivity is mainly limited by energy-degraded $\alpha$'s, emitted by surface contaminations on the crystals and on the copper supporting structure \cite{Bucci:2009fk,Alduino:2017qet}. High energy (four - six MeV) \al\ particles, in fact, lose only part of their energy in the crystal or surrounding materials and give rise to a continuum of events in the ROI. The amount of surface contamination is less than a few 10$^{-8}$ counts h$^{-1}$ keV$^{-1}$ cm$^{-2}$ and can not  be measured or screened with any standard technique.\\
As stated in section \ref{sec:limitation} the impossibility to disentangle  particles on the crystal surface or external to it and the lack of particle identification represents the actual limits of bolometers  in the search of \onu\ decay. \\
Given the enormous effort already devoted to surface treatment, it is unlikely that the required reduction in the background level can be achieved  by improving the radio-purity of the detector materials alone.
A major role to overcome this limit consists in  the development of new technologies for active background suppression. \\
This is the goal of the CUPID project (CUORE Upgrade  with Particle ID) \cite{Wang:2015raa,Wang:2015taa}  that aims at enhancing the  sensitivity for a bolometric experiment by two orders of magnitude increasing  the source mass and reducing the backgrounds using isotopically enriched bolometers with particle identification. 
The pursued approaches  are based on  different physical principles and different techniques and will be detailed in  section \ref{BAPI}. The reduction of the background induced by sources other than  surface ones are  reported  in section \ref{err}.

\subsection{Other second order weak processes}
In the \onu\ case discussed so far,  the decay was assumed to proceed to the ground state of the final nucleus. Given the short range of MeV electrons in a solid medium (few mm) the majority of the candidate events consists in a monochromatic energy release contained in a single crystal.  The decay  can also proceed to the excited states of the final nucleus. Although the longer predicted half life for these cases, they could be of experimental interest because the de-excitation \ga\ of the final nucleus give rise to a multi-crystals signature and therefore to a strongly reduced background. \\
However  the exchange of Majorana neutrinos between the two weak vertices can occur in transitions with a nuclear charge change of $\Delta$Z=-2  through $0\nu \beta^+ \beta^+$, $0\nu$EC$\beta^+$, $0\nu$ECEC decays modes, where EC is  the electron capture acronym. In the last mode no leptons are available to carry away the released energy and the decay can happen through  radiative  \cite{Sujkowski} or resonant decay \cite{Bernabeu}. 
The ECEC mode  is therefore typically suppressed but enhancements  (up to a factor 10$^6$) may happen when  there is degeneracy of the initial and final excited atomic states (resonance condition). The $\Delta$Z=-2 processes are interesting because they can provide an insight into the \onu\ mechanism since they are dominated by right-handed weak currents. \\
The most sensitive probe is represented by the \onu\ process to the ground state and  in the rest of the review only this decay is considered. Anyhow the potentialities of the future experiments are related to the background reduction capability and this aspect leads to significative improvements  for all modes, irrespective of their signature.

The \onu\ process is always accompanied by the \tnu\ decay which is allowed by the Standard Model since two neutrinos are emitted in the finale state.
This second order weak process has been observed for 11 nuclei \cite{Barabash:2015eza} and represent the rarest nuclear decays ever measured.
The experimental signature is weaker than the \onu\ mode and  consists in a broad spectrum from zero up the the \QBB\ value (neutrinos emitted at rest) and, as for the \onu\ case, most of events are contained in a single crystals. The extraction of the signal is therefore challenging because it must be disentangled from several background sources and  particle identification is not going to play a fundamental role since backgrounds are mainly \be/\ga\ i.e. same particle type of the signal.
On the other hand the limited size of actual cryogenic apparatuses imposes the use of bolometers enriched in the isotope under investigation, therefore the \tnu\ signal to background ratio will dramatically increase thus allowing  better measurements of the \tnu\ shape and intensity.
Finally it must be stressed the \tnu\ represent the irreducible background for the \onu\ search as discussed in section \ref{err}.


\section{Bolometers with Active Background Suppression}
\label{BAPI}

\subsection{Scintillating bolometers}
\label{sb}
 The use of heat-scintillation hybrid bolometers for the \onu\ search was proposed in 2005 \cite{Pirro:2005ar}.\\
 In a scintillating bolometer \cite{nsvecchioarticoloCaF2,Pirro:2005ar,Giuliani2012} a small fraction (between one per cent and one per mill) of the released energy $E$ is converted into scintillation light. This eventually escapes the crystal and is absorbed by a thin bolometer working as light detector.\\
The ratio between the two signals (scintillation/heat) depends on the particle type; \be/\ga\ particles have a light yield (LY) which is typically different from the LY of \al\ interactions or neutrons that are quenched. Consequently, the  dual readout allows particle identification.
 
In addition, the freedom in the choice of the absorber provides the unique opportunity of selecting the \onu\ isotopes with high \QBB \cite{Arnaboldi:2010tt,Gironi:2009ay, Arnaboldi:2010jx}. \\
This is a very important aspect.  The most prominent natural high-energy \ga s, induced by the  \TL\ decay, are distributed up to 2615 keV  while  only rare \ga\ decays coming from  the $^{214}$Bi  decay populate the above region and up to 3270 keV. 
 Selecting a \onu\ candidate with a \QBB\ larger then 2615 keV will leads to a \ga\ background reduction in the ROI by about one order of magnitude as can be inferred  from the \ga\ spectrum measured at the Laboratori Nazionali del Gran Sasso \cite{Bucci:2009fk}.\\
With a proper absorber choice, scintillating bolometers can simultaneously get rid of both \al\ induced background and the most intense natural \ga\  radioactivity.\\
In the last ten  years several scintillating bolometers were operated  with remarkable results using as \onu\ emitters: $^{82}$Se(\QBB =2998 keV\cite{wang2016}), $^{100}$Mo(\QBB=3034 keV\cite{wang2016}), $^{116}$Cd (\QBB=2813 keV\cite{wang2016}). The effort was devoted not only to the development of the light detector technology but also to the production of very pure enriched crystals (see section \ref{err}).

Three  small scale pilot experiments using scintillating and isotopically enriched crystals are (or close to) being  operated as final demonstrator in view of a next-generation experiment. CUPID-0 \cite{Artusa:2016maw}, formerly LUCIFER \cite{Beeman:2013sba}, is currently running using an array of 24 enriched Zn$^{82}$Se crystals; LUMINEU/CUPID-0-Mo \cite{Armengaud:2016dqg, Armengaud:2017hit} will start data taking in January 2018 with an array of 20 Li$_2$$^{100}$MoO$_4$ crystals; AMORE \cite{Kim:2015pua} is operating 5 $^{48-dep}$Ca$^{100}$MoO$_4$ bolometers but it foresees the use  of other molybdates such as Zn$^{100}$MoO$_4$ or Li$_2$$^{100}$MoO$_4$ for the final experiment.
The first two mentioned experiments are part of the CUPID R\&D.  No demonstrators for  $^{116}$CdWO$_4$  bolometers are on going or  planned despite the good results obtained \cite{Danevich:2016xcm}; hence only results for $^{100}$Mo and $^{82}$Se are discussed.

One of the greatest  advantages of scintillating bolometers  is that the amount of collected light is high enough to be recorded using a germanium slab operated as bolometer and equipped with a standard  Neutron Transmutation Doped (NTD) germanium thermistor. 
NTDs are heavily doped semiconductors,  obtained  by thermal neutron irradiation \cite{Hallerf, Haller} and with impurity concentrations slightly below the metal-insulator transition. In this variable-range-hopping regime, their resistivity depends exponentially on the temperature. They are high impedance devices (1-100 $M\Omega$), read out in  constant current biasing mode and matched to room temperature JFET amplifiers.
They are commonly used for the heat channel readout. The possibility to use the same sensor with minimal modification also for the light channel represents a clear advantage in terms of reliability, robustness and impact on the cryogenic infrastructure in view of a reuse of the existing CUORE infrastructure for a future experiment.   
 Those light detectors have been extensively characterized \cite{Beeman:2013zva} and the technology could be considered already mature. Light detectors show a baseline RMS noise of  hundreds of eV and allow particle identification  even in the case of the worst scintillators (LY$\sim$1 keV/MeV) \cite{Artusa:2016maw,Armengaud:2017hit,Beeman:2013zva}.
 
 The ratio of the light signal associated to an \al\ interaction (QF)  to  \be/\ga\ on for events with the same heat energy release is defined Quenching Factor (QF). It is of the order of 0.2 for most of the studied compounds with the only exception of the ZnSe which exhibits a QF of 3-6 \cite{Arnaboldi:2010jx,Beeman:2013vda,Beeman:2013sba,Artusa:2016maw}.

The discrimination capability of a scintillating bolometer is usually  parametrized  by a quantity, the Discrimination Power (DP), defined as 
\begin{equation}
\mathrm{DP} = \left|\mu_{\beta /\gamma}-\mu_{\alpha}\right|/\sqrt{\sigma_{\beta/\gamma}^2+\sigma_\alpha^2},
\label{dp}
\end{equation}

where $\mu$ and $\sigma$ denote the average value and the standard deviation of the $\alpha$ or  \be/\ga\  distributions respectively and quantities are computed at \QBB\ since they might depend on the energy. The DP can be the light/heat ratio but can also be applied to any  pulse shape variable.
 
ZnSe and some molybdates, in fact, show a peculiar feature: the thermal pulse induced by an \al\ particle has a slightly faster decay time than that induced by \be/\ga\ interactions \cite{Gironi:2009ay,Beeman:2012jd,Armengaud:2017hit,Artusa:2016maw,Gironi:2010hs,Beeman:2013vda,Arnaboldi:2010jx,Arnaboldi:2010gj,Beeman:2012gg,ZnMo4poda,Casali:2017zvs,Kim:2017xrs,Kim:2015pua}.

\begin{figure}[htb]
\centerline{\includegraphics[width=8.0cm]{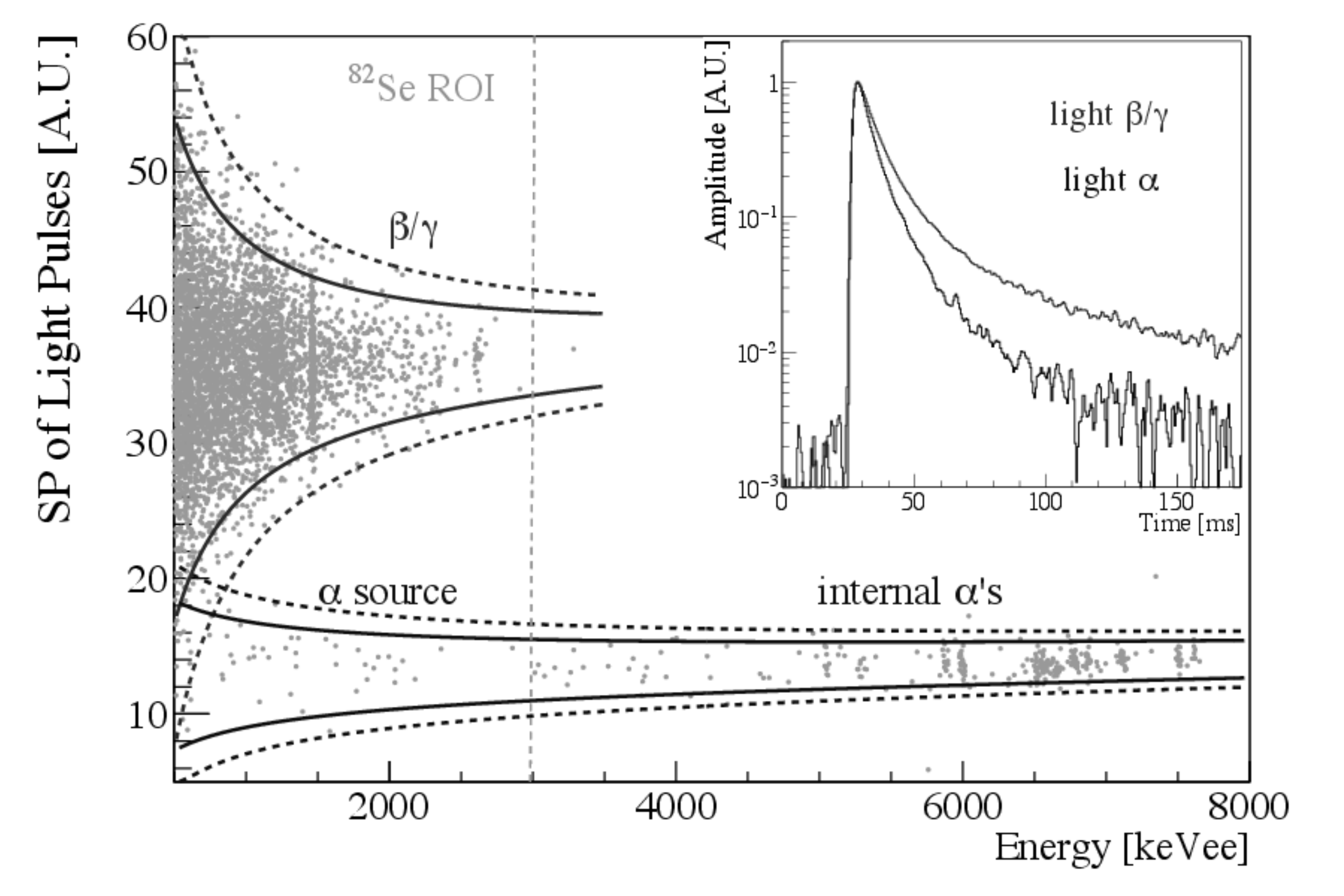}}
\caption{Shape parameter of a light detector as a function of the energy released in a Zn$^{82}$Se bolometer. The energy scale is calibrated on \be/\ga\ and thus referred as keVee. The  lines indicate the 2$\sigma$ (continuous) and 3$\sigma$ (dashed) \be/\ga\ and \al\ bands. \al\ events produced by a smeared Sm source (below 3MeVee) and by contaminations of the crystal bulk (peaks above 5MeVee) can be easily rejected, in particular in the region of interest for the $^{82}$Se \onu (dashed vertical green line).  Inset: time development of light pulses produced by \be/\ga\ and a \al\ interactions with energy of about 2.6 MeV. A DP of 12 is  obtained at the $^{82}$Se \QBB. Figure adapted from Ref. \citen{Artusa:2016maw} with kind permission of the European Physical Journal (EPJ).}
\label{znse}
\end{figure}
\begin{figure}[hbt]
\centerline{\includegraphics[width=8.0cm]{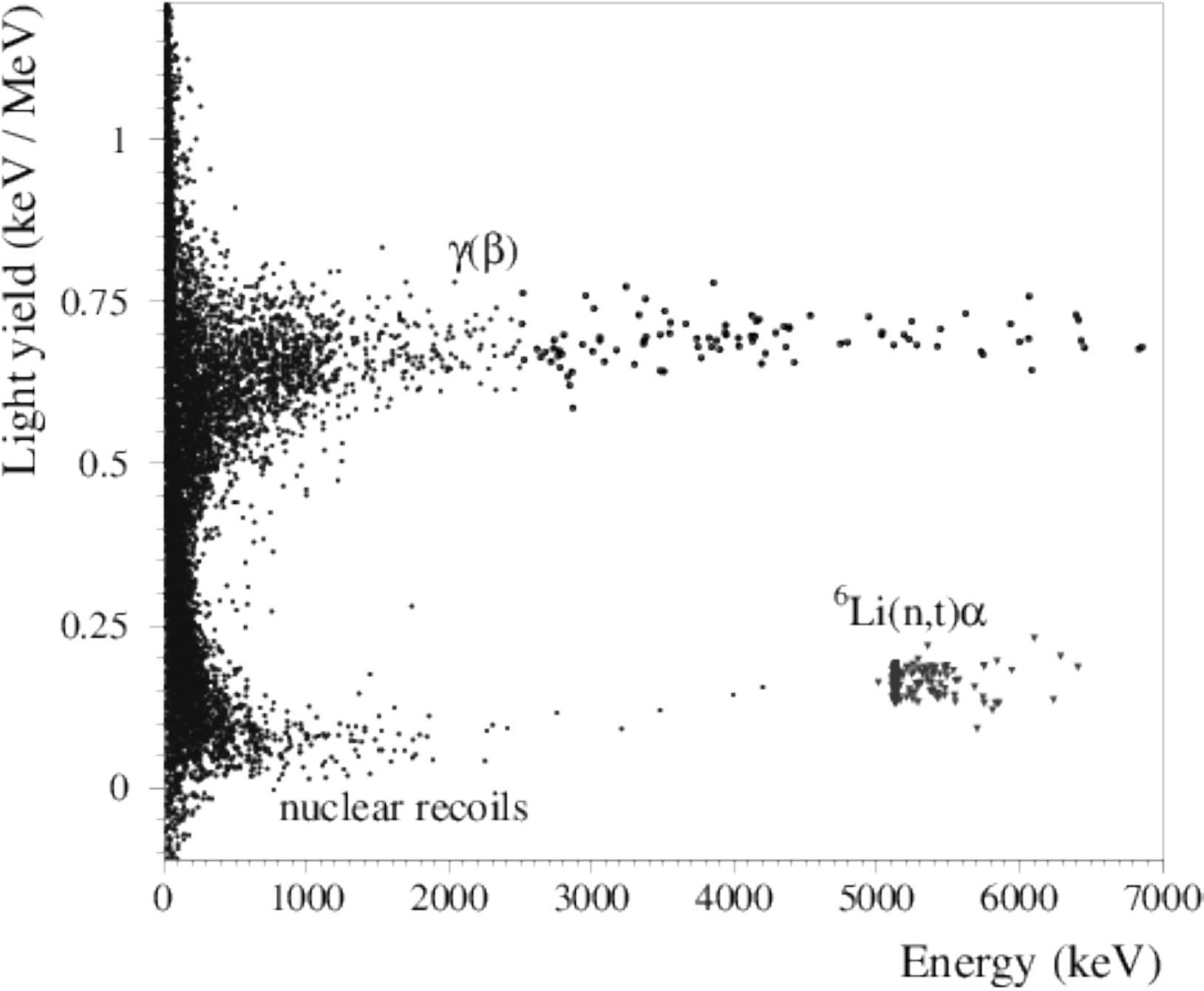}}
\caption{Light Yield-vs-heat scatter-plot obtained with  AmBe neutron source  with a 151 g Li$_2$MoO$_4$ scintillating bolometer.  A clear separation between  \be/\ga\ and  \al\ interactions is visible. The \be/\ga\  band exceeds the natural $^{208}$Tl end-point because of the prompt de-excitation \ga s following $^9$Be(\al,n)$^{12}$C$^{\star}$ reaction. The cluster of events in the \al\ region is caused by the reaction $^6$Li(n,t)\al. Figure adapted from Ref. \citen{Armengaud:2017hit} with kind permission of the European Physical Journal (EPJ).}
\label{limo}
\end{figure}

This is a  tiny (few percent) effect that allows to discern the nature of the interacting particle without light detection thus greatly simplifying the detector assembly and readout. This effect could be ascribed to the  long scintillation decay time (of the order of hundreds microseconds)  and  the high  percentage of non-radiative de-excitations of the scintillation channels, that produce delayed phonons \cite{GironiPSD}. However a very good signal-to-noise ratio for all the channels is required because of the smallness of the effect and  the discrimination power was not always reproducible in different experimental measurements. Further developments are necessary before considering a reliable technology.\\
In the case of  ZnSe bolometers,  a pulse shape difference, more discriminating than the one on the heat bolometer, is seen on the light channel \cite{Beeman:2013vda}. This is currently used  by the CUPID-0 collaboration \cite{Artusa:2016maw} to avoid the leakage of the \al\ band of the LY into the \be/\ga\ band as observed  in the light-vs-heat scatter plot \cite{Beeman:2013vda}. 
Examples of distributions of discriminating variables are reported in Fig. \ref{znse} and Fig. \ref{limo}. \\
The requirements for a bolometric experiment  to assess the IH hierarchy mass region imply a  background level in the ROI of $10^{-4}$ \cpyI  \cite{Beeman:2011bg}. This requires  a  rejection factor on \al s better than 99.9\%  while preserving a signal efficiency greater then 90\%  \cite{Beeman:2011bg}. A DP of 3.1 or greater is necessary to satisfy this criterion. Table \ref{tab:PID} reports the the \be/\ga\ LY, QF and DP for all the compounds for which a pilot experiment is running or in construction phase; all of them exhibit a DP exceeding 9, much above  the requested threshold.

\begin{table}[hbt]
\tbl{The table reports the light yield (LY), quenching factor (QF), and discrimination power (DP) for the enriched scintillating bolometer demonstrators quotet in the text.}
{\begin{tabular}{@{}lllll@{}}
\toprule
Bolometer									& $LY_{\beta/\gamma}$(keV/MeV)	& $QF$ & $DP$	& Refs.  \\
\colrule
\hline
ZnSe											& 2.6--6.4		& 3--4.6				& 9--17	& \cite{Beeman:2013vda,Arnaboldi:2010jx}  \\
Zn$^{82}$Se								& 3.3--5.2		& 2.7						& 10--12	& \cite{Artusa:2016maw}  \\
\hline
$^{40}$Ca$^{100}$MoO$_4$	& n.a.				& 0.19--0.33		& 9--11	& \cite{kim-ieee}  \\	
\hline
Li$_2$$^{100}$MoO$_4$			& 0.73--0.77	& 0.15--0.22		& 12-18	& \cite{Armengaud:2017hit,Poda:2017jnl}  \\
\botrule
\end{tabular} 
\label{tab:PID}}
\end{table}

\subsection{Cherenkov light in TeO$_2$ bolometers}
\label{nonsb}

The TeO$_2$ bolometers used by the CUORE collaboration do not show any significant scintillation. A tiny light signal was observed in 2004 \cite{coron} and seems to find confirmation recently \cite{Berge:2017nys}. In any case  the light yield  is low ($\sim$20 eV) and  negligible compared to another and more important process that takes place: the emission of light through the Cherenkov effect. 

\begin{figure}[hbt]
\centerline{\includegraphics[width=9cm]{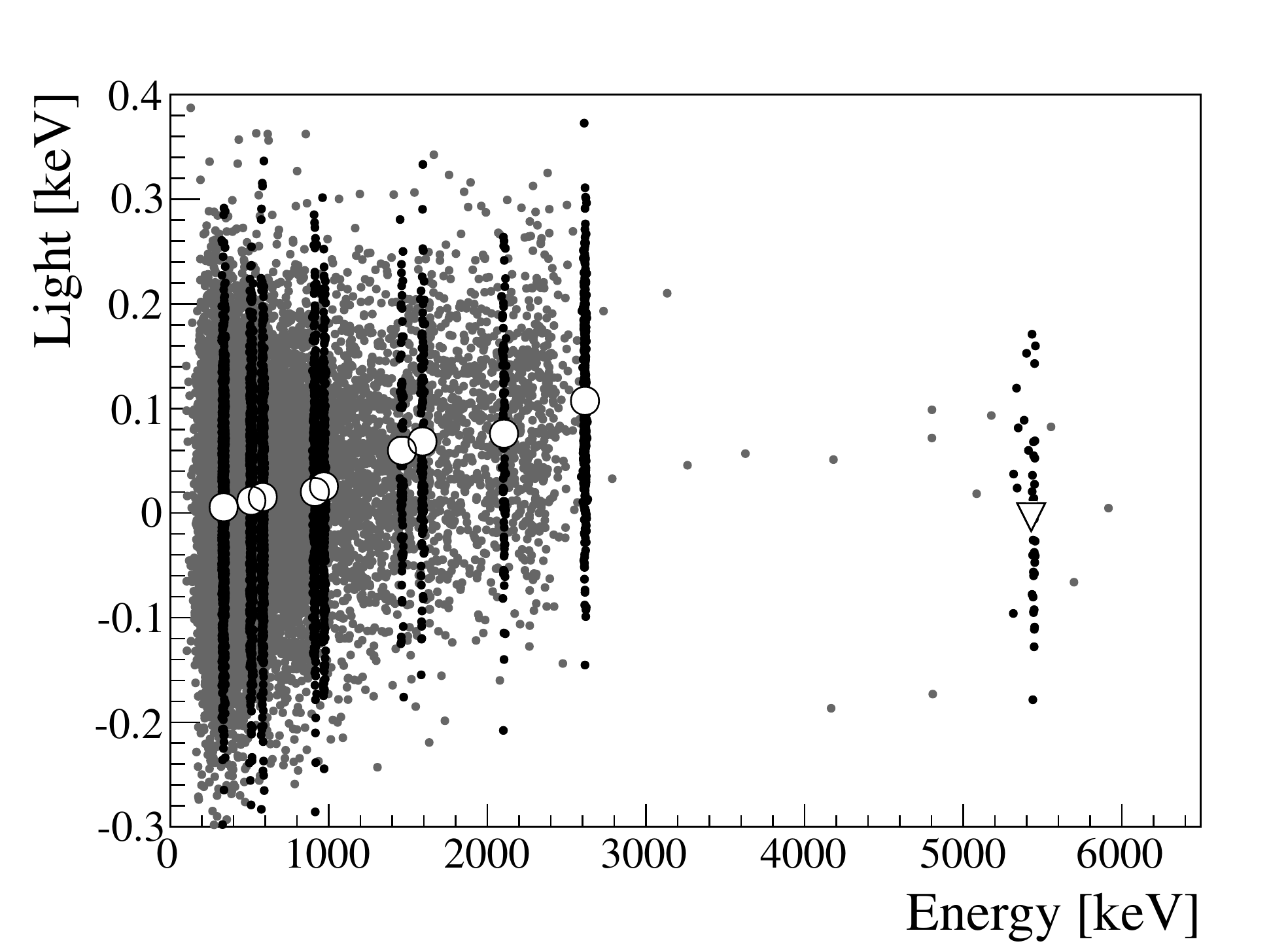}}
\caption{Detected light versus calibrated heat in a CUORE-like TeO$_2$ bolometer read out with a CUPID-0 NTD Ge thermometer. The mean light is clearly energy dependent for the \be/\ga\ peaks (circles below 3 MeV) and compatible with zero for the \al\ decay of the $^{210}$Po (triangle).
TeO$_2$ does not scintillate, however Cherenkov light is produced by \be/\ga\ interactions (circles) and not by \al\ ones (triangle). 
Figure adapted from Ref. \citen{Casali:2014vvt} with kind permission of the European Physical Journal (EPJ).
}
\label{Cherenkov}
\end{figure}

In 2010 the use of the Cherenkov light in TeO$_2$  bolometers was suggested as a tool for particle identification \cite{TabarellideFatis:2009zz}.
The threshold for Cherenkov emission  in TeO$_2$ is around 50 keV for electrons and around 400 MeV for \al s. At the energy scale of interest for \onu , the signal electrons emit light while \al\ particles do not.
Several tests were done on small \cite{Bellini:2012rc,Willers:2014eoa,Gironi:2016nae} and large crystals \cite{Beeman:2011yc,Schaffner:2014caa,Casali:2014vvt,Casali:2015gya,Artusa:2016mat, Berge:2017nys} to characterize the discrimination power.\\
 The challenge of this method is the detection of the extremely small amount of light emitted by electrons at the $^{130}$Te \onu\ energy (\QBB $\sim$2.5 MeV) that is of the order of 100 eV \cite{Casali:2013bva,Casali:2014vvt,Casali:2016luq}, i.e.  comparable to the noise resolution of  NTD-based standard light detectors used in scintillating bolometers (see Fig. \ref{Cherenkov}). A signal-to-noise ratio greater  than  5 is needed to reach \al/(\be/\ga) separation allowing for a 99.9\% rejection of the \al\ background \cite{Casali:2014vvt}.
Attempts to increase the light collection \cite{Casali:2014vvt} do not lead to significative results, this implies that  a light detector technology with a noise level below 20 eV RMS is mandatory.
Furthermore the light detectors must be robust and reproducible in view of a ton-scale experiment with about 1000 bolometers,  made of radio-pure materials and  possibly have a multiplexed readout to avoid a large heat-load on the cryogenic apparatus. Finally the light detector must have an active area comparable to the top  bolometer face (about 20 cm$^2$) in order to maximize the light collection. 
A carefully optimized Ge bolometer with a NTD-Ge sensor  \cite{Coron2004}  achieved the required performance in terms of resolution but reproducibility and robustness are far to be demonstrated.

The next three sections are dedicated to the status and perspectives of the  light detector technologies under development.

\subsubsection{Transition edge sensors and metallic magnetic calorimeters}
\label{tes}
\begin{figure}[h]
\centerline{\includegraphics[width=9cm]{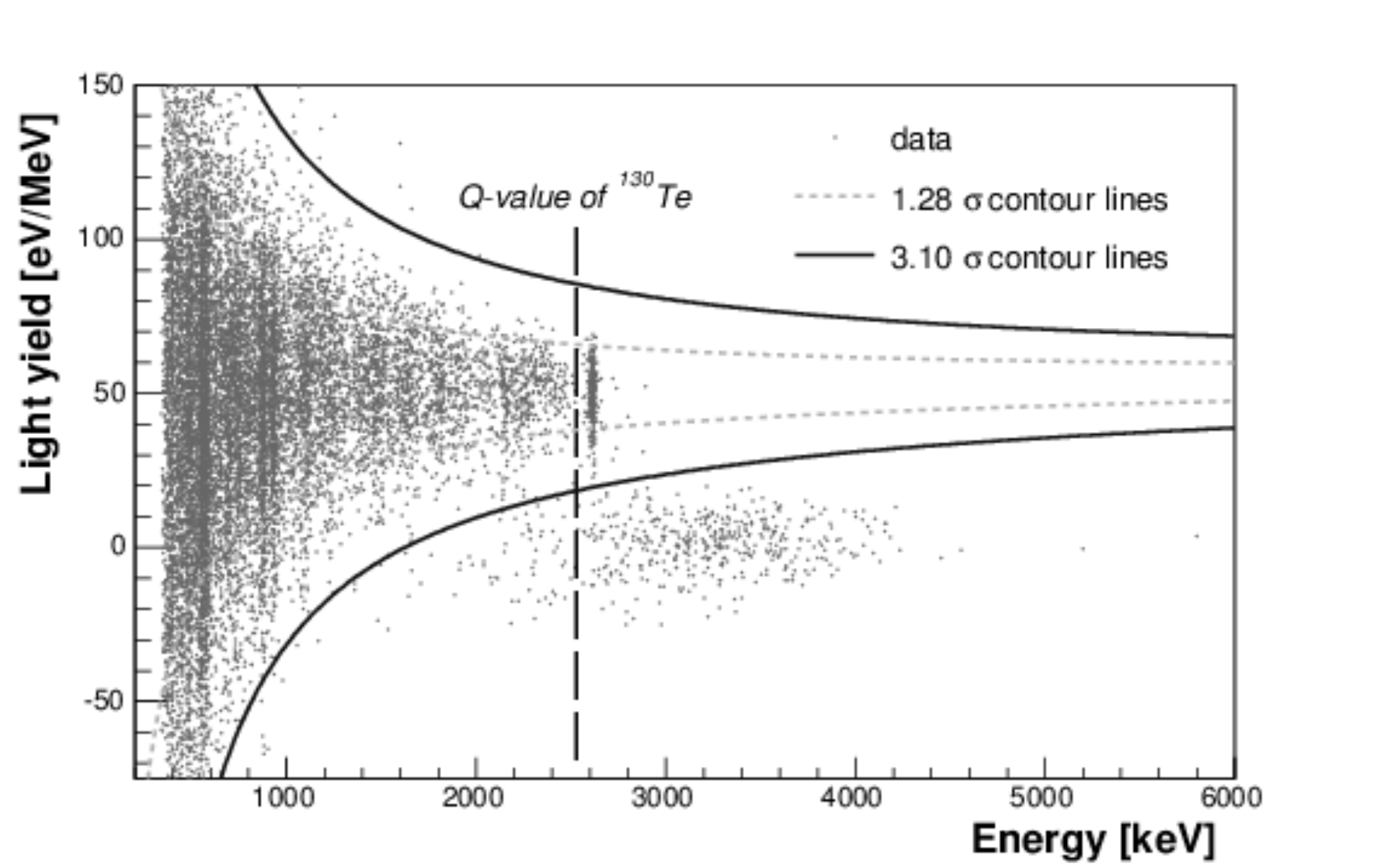}}
\caption{Background data of the TeO$_2$ bolometer is shown in the light yield-energy plane. Light yield obtained with a massive (285 g)  TeO$_2$ bolometer and TES-equipped Silicon on Sapphire light detector. Two distributions can be noted: a band due to \be/\ga\ interactions as well as the less populated band at zero light yield due to \al\ particles from a degraded \al\ source. The bands which indicate the region expected for \be/\ga\ events are shown in form of central probability bands. The dotted lines are $\pm 1.28 \sigma$ contours whereas the solid lines are $\pm 3 \sigma$ contours, thus 99.8\% of all \be/\ga\ events are expected to be contained within the two solid contour lines. A DP of 3.7 is achieved. The \al\ particle distribution appears at a light yield of zero, separated from the populated \be/\ga\ band. The dashed vertical line indicates the Q-value of $^{130}$Te of 2530 keV. 
Figure adapted from Ref. \citen{Schaffner:2014caa} with  permission from Elsevier.
}
\label{Che_tes}
\end{figure}

A technology, able to reach the desired energy resolution  on a large area light detector, does exist and is currently implemented by the CRESST dark matter experiment \cite{Angloher:2011uu,Angloher:2014myn,Angloher:2015ewa}.
A half mm thick sapphire disc, with a one micron layer of silicon on it, is equipped with thin tungsten Transition Edge Sensor (TES)  coupled to
an aluminum absorber.
A TES sensor is a resistive device that operates at the critical temperature $T_c$ of the superconductor so that the resistivity changes sharply from zero to a  finite value in a very narrow temperature interval. TESs are biased at a constant voltage and their low impedance (in the  few
m$\Omega$ - $\Omega$ range) imposes the use of Superconducting Quantum Interference Device  (SQUID) amplifiers. They are intrinsically fast devices with a bandwidth of MHz or more.  This offers,  in addition to the excellent energy resolution, two advantages: the pulse shape sensitivity is significantly
improved  and time resolution better than one ms can be achieved.
A CRESST light detector, coupled to a CUORE-style bolometer,  demonstrated  an event-by-event  basis \al/(\be/\ga) separation \cite{Schaffner:2014caa} (see Fig. \ref{Che_tes}) but scaling of the technology to a thousand detectors requires  a dedicated development on the reproducibility of the technology (e.g. uniformity of transition temperature  across many channels) at temperatures of about ten mK and on the readout multiplexing capability  to reduce the wiring complexity and the heat load. Solutions exist in the astrophysics community \cite{Nucciotti:2015rsl} but the portability to the \onu\ research field is not trivial. 
They are based on the use of RF-SQUIDs  coupled to  superconducting coplanar waveguide (CPW) GHz resonators and homodyne detection. Tuning the resonators at different frequencies, it is possibile to multiplex several  RF carriers (see section \ref{mkid}).
This approach, called Microwave Multiplexing ($\mu$MUX), has been demonstrated for two channels \cite{Noroozian} and  is quickly developing \cite{Dicker} but it has been shown to work up to now only for compact arrays of micro-calorimeters with mass much less than one milligram.

Other TES implementations are under study  in the \onu\ community \cite{Wang:2015taa}. The first  aims  at reaching a  lower $T_c$ making use of the proximity effect in  bilayer films of superconductor and normal conductor (e.g. Ir-Au, Ir-Pt, Mo-Au, etc).  The second makes use of  NbSi, which is a superconductor for an appropriate stoichiometric ratio with an intrinsic high resistivity in the normal state (1-5~M$\Omega$). 
This would allow the use of the same conventional electronics, based on JFETs, as for NTDs when they are operated within the transition. This solution does not provide all the advantages related to the low-impedance TESs, but it is possible to get a temperature sensitivity up to ten times higher of that achieved by NTDs keeping the same front-end electronics, and so with a minimal impact on the  readout structure. \\

Another class of sensors, which share similar characteristics with TESs, is represented by Metallic Magnetic Calorimeters (MMC).
They  base their principle  on the strong temperature dependence of the magnetization in paramagnetic sensors and are typically made of Au:Er.
A  variation of the magnetic moment  can be read out with high sensitivity using meander-shaped thin-film pickup coils and SQUID magnetometers. This effect, already exploited with outstanding results in X-ray spectroscopy \cite{porst}, can be used to develop exceptionally sensitive thermometers. They are very fast sensors (rise-time below 50~$\mu$s) and can reach an energy resolution better than ten eV. Because of these two features, their multiplexed readout is even more demanding than that of TESs and the only feasible approach is  $\mu$MUX \cite{kempfs}. MMCs are  adopted by the AMORE collaboration \cite{Kim:2017xrs} although the amount of scintillation light produced  by the $^{48-dep}$Ca$^{100}$MoO$_4$ crystal does not require a very sensitive light detector. No multiplexing readout is applied in this case.

\subsubsection{Kinetic inductance detectors}
\label{mkid}
The working principle of a Kinetic Inductance Detectors (KID)  is based on the change of its kinetic inductance  when the density of  Cooper pairs is modified \cite{day}.
In superconducting materials the Cooper pairs, characterized by a binding energy smaller than 1$\,$meV, move through the lattice without scattering. 
If a RF electromagnetic field is applied, the pairs oscillate and acquire a kinetic inductance. The inductor is inserted into a high quality factor ($Q>10^3$) RLC circuit giving rise to a resonator with a resonant frequency  $f_0=1/2\pi\sqrt{LC}$. An energy release $E$, able to break Coopers pairs into quasi-particles, changes the kinetic inductance and thus the transfer function and could be inferred by  the variations in phase and amplitude of the transmitted signal.
 KID detectors are a leading technology in astroparticle physics  \cite{Monfardini:2011yh,Mazin:2013wvi} and their use in the \onu\ field was proposed by the CALDER project \cite{Battistelli:2015vha}.\\
Their strengths are: several  KIDs  can be coupled to the same feedline and can be multiplexed by making them resonate at slightly different frequencies since $f_0$ can be easily changed by slightly modifying the layout of the capacitor and/or inductor of the circuit; the readout electronics is quite simple and operated at room temperature, with the exception  for a low noise cryogenic amplifier; performances do not depend critically on the working temperature, provided it  is well below the critical temperature  of the superconductor.\\
The main drawback is that dimensions must be smaller than the wavelength of the excitation signal, so that the current in the inductor is uniform and the signal does not depend on the position of the energy release. Their size is limited to a few mm$^2$, by the optimal range (1-4 GHz) of already available electronics and  the number of resonators that can be coupled to the same line. 
To reach a large surface light detector, KIDs are deposited on silicon substrate as in CRESST light detectors~\cite{Angloher:2011uu}. 
Photons impinging on the back side of the chip  produce ballistic phonons which scatter through the substrate and  reach the KID on the opposite surface \cite{Moore:2012au,Swenson:2010yf}. To compensate the efficiency loss with respect to direct absorption, a few KIDs per light detector are needed. 
In the last 3 year the CALDER \cite{Battistelli:2015vha} project developed and tested several KID detectors using Aluminum as superconducting material \cite{Cardani:2015tqa,Casali:2015bhk,Colantoni:2016alu,Martinez:2016rks,Vignati:2016adb,Colantoni:2016tpk,Bellini:2016lgg,Casali:2017yro}.
With a 4  mm$^2$ single KID resonator on a 2x2 cm$^2$ substrate, an energy resolution of about 80 eV has been achieved \cite{Bellini:2016lgg}.

The energy resolution of KIDs scales as T$_c/\sqrt{QL}$\cite{Zmui, McCammon}. Too boost it to the desired level, different superconductors with optimized  T$_c$ and  L were investigated. The first large area (25 cm$^2$) detector made with  Al+Ti+Al KID is being measured  and results will be published soon.

\subsubsection{Neganov-Trofimov-Luke effect}
\label{nl}
If a light detector  comprises a semiconductor substrate, its baseline noise resolution could be enhanced exploiting the Neganov-Trofimov-Luke (NFL) effect \cite{luke,neganov}.
An electric field  applied to the device, in fact, accelerates electron and holes generated by an energy release $E$ inside the detector itself. The work done by the field on the charges produces  an enhancement of the thermal signal recorded by the thermometer attached to the semiconductor wafer. 
The total energy $E_t$ dissipated is 
\begin{equation}
E_{t} =  E(1+\frac{eV}{w}), 
\label{NL}
\end{equation}
where e is the electron charge,  V is the applied drift voltage across the electrodes, and $w$ is the mean energy needed to create an electron-hole pair. The amplification is independent of any other source of noise and allows to lower the baseline noise resolution and  decrease the energy threshold. This mechanism is well known and used in dark matter searches \cite{stark,Isaila:2011kp,cdms,edel} and in the last two years has been successfully applied to detect the Cherenkov light in TeO$_2$ bolometers with different devices with both germanium and silicon absorbers.\cite{Willers:2014eoa,Casali:2015gya, Artusa:2016mat,Gironi:2016nae,Biassoni:2015eij}.  \\
 Recently a complete event-by-event \al/(\be/\ga) separation in a full-size TeO$_2$ CUORE bolometer coupled to a NTD-based germanium light detector with NTL amplification has been achieved \cite{Berge:2017nys}.  In this case the electrodes, a set of concentric Al rings on a side,  generate an electric field parallel to the surface that allows  to decrease the charge trapping probability thanks to the short path length of the charges to the electrodes. This represent a fundamental result  in view of an application in CUPID \cite{Wang:2015raa} since it could be adopted with minimal modification of the entire readout with the respect to the one actually in use in the CUORE experiment.

Devices with silicon absorbers and TES  \cite{Willers:2014eoa}  and NTD \cite{Gironi:2016nae,Biassoni:2015eij} sensors were also developed. 
In the case of NTD sensors, the advantages compared to germanium absorber hinge on the wider  range of processing technologies for silicon, that potentially allows the integration of thermal sensors and mechanical suspension structures. A further advantage of silicon over germanium is the fact that specific heat of silicon is a factor four smaller than germanium, opening the possibility of building substantially larger detectors without compromising the signal amplitude which is inversely proportional to the heat capacity of the device. Promising results have been obtained with 2x2 cm$^2$ area detectors\cite{Biassoni:2015eij} and  first  5x5 cm$^2$ sample are under measurement.  
Plans to use the NFL effect with  KID sensors for single photon counting are also under investigation.

\subsection{Surface sensitive bolometers}
\label{shape}
Tagging surface events is difficult, as bolometers are fully sensitive device in the volume and often present a single response to any type of fast energy deposition, irrespective of its nature and location.
Even when \al\ background could be reduced at the desired level, a non negligible contribution could be represented by single \be\ particles emitted in decays of $^{214}$Bi as well as by $^{210,208}$Tl decays that emit electron and  \ga s in coincidence producing a single event which escapes delayed coincidences tagging (see section \ref{err}).
 
An alternative approach, to those aiming of hybrid bolometers, consists into  achieving  impact-point sensitivity making use of superconducting  Al films (about ten $\mu$m) deposited on the detector surface which can modify the signal shape of surface events \cite{Schnagl:2000}.
The physical principle is the following: athermal phonons generated by a particle that releases its energy within a few mm from the surface (\al\ or \be s)  break Cooper pairs in the superconducting film and produce quasi-particles, which have  a long lifetime (on the order of milliseconds) in high purity aluminum. The quasi-particles  recombination produces ballistic phonons, that will add a delayed component to the leading edge of the signal read out by the sensor on the main bolometric absorber. 
For bulk events, instead, the athermal phonon population reaching the Al film is more degraded in energy and less efficient in producing quasi-particles.  Surface events will have therefore  longer rise-time compared to bulk interactions. This mechanism has been evidenced in Ref. \citen{Oli2008} and the proof of principle of this technique for \onu\  decay detector has been demonstrated in a TeO$_2$ bolometer with deposited Al film and using fast phonon sensors based on NbSi films, with rise times on the order of one ms \cite{Nones:2012}. 

Unfortunately, the current NbSi sensor technology is  unsuitable for \onu\ search because the important component of athermal phonons in the signal 
induces position dependent amplitude and thus deteriorate the energy resolution. The recently  ERC-approved project CROSS \cite{Giuliani:2017} 
aims at achieving surface-to-bulk signal separation  with the use of NTD sensors, i.e. with a heat pulse rise-time on the order of tens of milliseconds. This might be possible as the excellent signal-to-noise ratio characterizing the typical CUORE readout has the potential to highlight even tiny pulse-shape differences.  This technique could be applied also to scintillating bolometers since once the surface $\alpha$ background is rejected, the dominant contribution arises from surface $\beta$s contaminations \cite{Artusa:2014wnl}.

\section{Environmental  Radioactivity Reduction}
\label{err}

\begin{figure}[h]
\centerline{\includegraphics[width=6.5 cm]{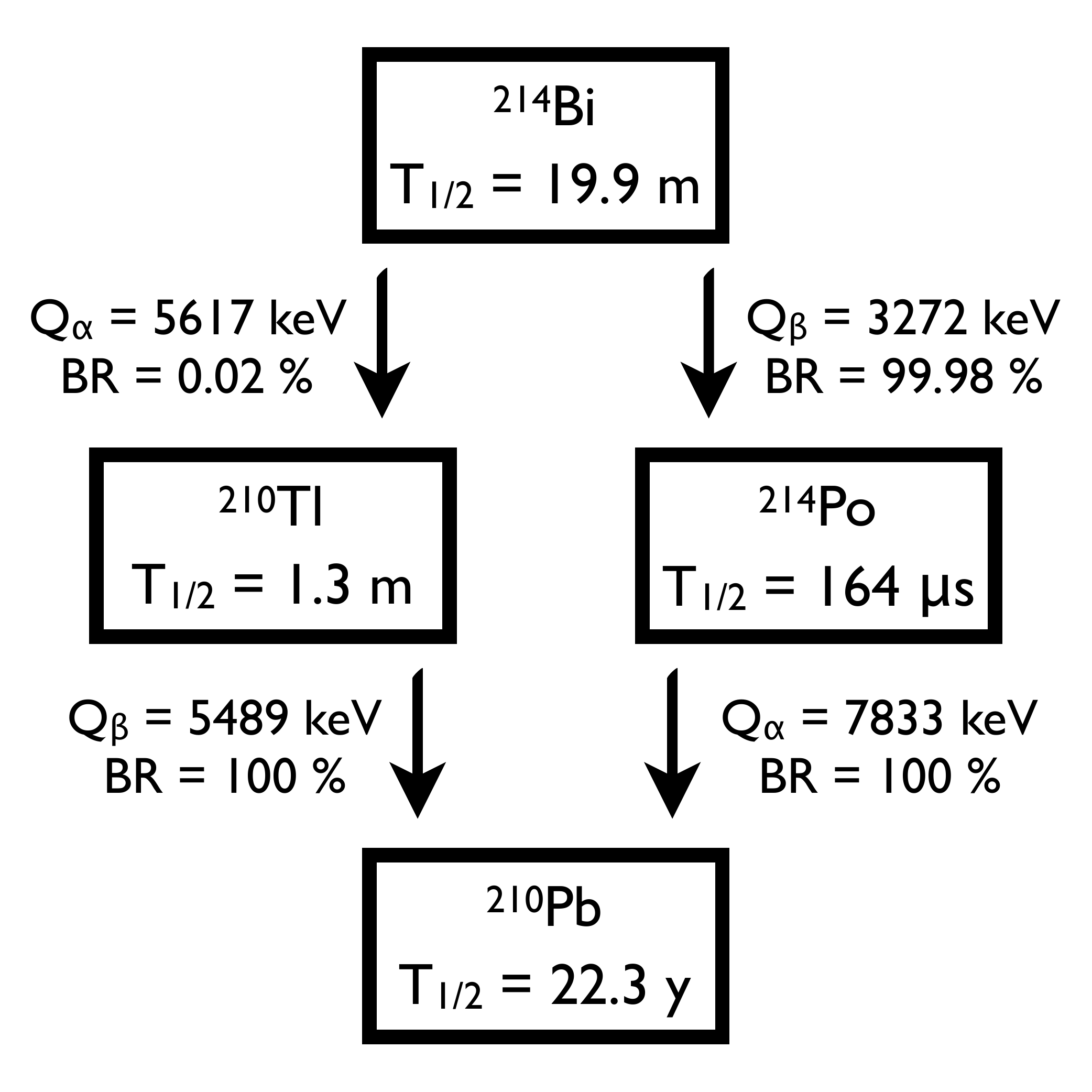}\includegraphics[width=6.5cm]{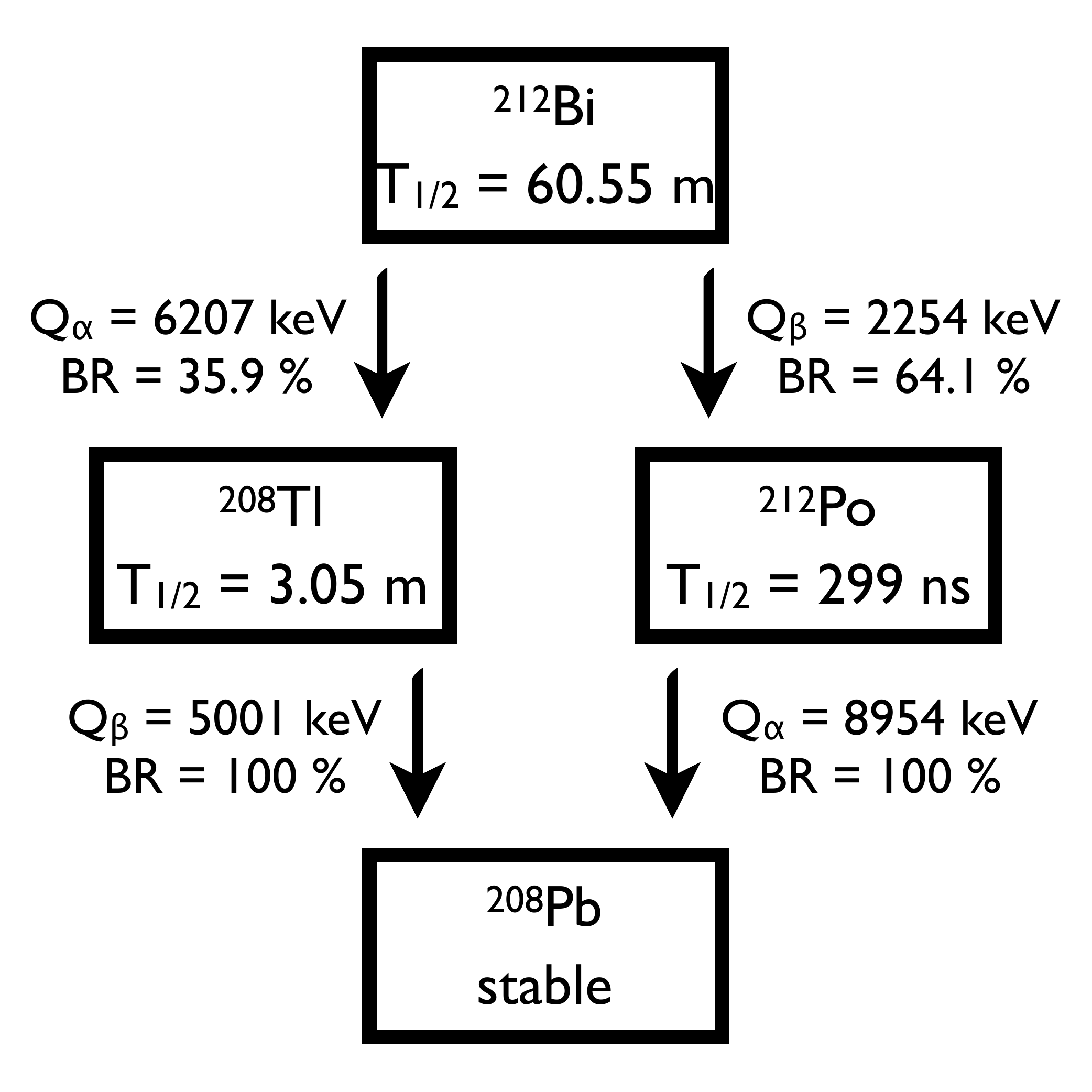}}
\caption{Left: decay scheme of $^{214}$Bi. The short life of its daughter, $^{214}$Po, causes the pile-p of the \be\ emitted by $^{214}$Bi and the \al\ particle produced by $^{214}$Po. Right: decay scheme of $^{208}$Tl. The delayed coincidence with the \al\ particle produced by its mother, $^{212}$Bi, allows to suppress this background source. Illustration  is a courtesy of Laura Cardani. 
}
\label{BiPo}
\end{figure}

The particle identification techniques, on which detector developments are mainly focused, aim at reducing to negligible levels the effect of surface contaminations of detector materials that represents the dominant background in  CUORE. 
However, the reduction of surface contamination effects can't by itself ensure the achievement of a two orders of magnitude background reduction as foreseen in CUPID. 
Other different sources, as bulk contaminations of crystals, copper supporting structure, lead shields, and the -small parts- as glue, bonding wires or readout cables and pads,  can contribute to the ROI counting rate at levels of $\sim 10^{-3}$ \cpyI \cite{Artusa:2014wnl}.\\
All of this results in serious restrictions in the use of materials. Stringent purification protocols for crystal production must be developed and all the materials close to the detectors have  to  be fabricated from radio-pure materials and  assembled  in radon free environment with dedicated radio-pure tools. A special attention has also to be paid to avoid  cosmogenic activation.
An exhaustive list of low background techniques exploited in this research field can be found in  Ref. \citen{Poda:2017jnl}.\\
The bulk activity of the crystal absorber must be controlled to a level such to not spoil the background index the ROI. 
Internal \al\ decays from U/Th chains can not contribute to background since they give rise to sharp peaks with energies Q$_{\alpha} >$ 4 MeV, i.e. far above the ROI.
Internal $\beta$ decays with Q$_{\beta} >$ 3 MeV  could represent instead a worrisome background due to their continuum spectrum.  They are generated by $^{214}$Bi ($^{238}$U chain) and its daughter $^{210}$Tl and by $^{208}$Tl ($^{232}$Th chain) as reported in the scheme of Fig. \ref{BiPo}.
Given the slowness of the bolometric response the $^{214}$Bi $\beta$ decay  is followed by the  fast $^{214}$Po \al\ decay  and their energy is summed up far from the ROI (Bi-Po events). The $^{208}$Tl and $^{210}$Tl  decays instead to stable  $^{208}$Pb  and $^{210}$Pb respectively; anyway  they could be tagged by delayed coincidence with the primary \al\ decay. In order to limit the dead time introduced with this technique (\be\ half life of the order of minutes) the total activity of U/Th contamination in the bulk must be kept at the level of the $\mu$Bq/kg.\\
 Another important aspect which must be taken into account is that the increase of the sensitive mass demands the production of high-quality, radio-pure enriched crystals.  The difficulty in operating cryogenic systems with an experimental value larger  than the existing CUORE one makes the  isotope enrichment the only viable way for enhancing the mass of the isotope of interest. The use of enriched material has consequences on the purification/crystallization chain. The enriched material   could have residual chemical impurities which may require additional purification stages to get high-quality crystals. Enrichment is generally done by gas centrifugation in  facilities used to separate different isotopes  without special radio-purity concern and therefore an additional purification is required.
Furthermore the enriched material is expensive, and the growth procedure must be adapted in order to reduce as much as possible the irrecoverable losses of the initial charge. Crystal bulk contaminations \cite{Poda:2017jnl}  are actually at the level of few to ten  $\mu$Bq/kg and are approaching the target value to reduce the internal  background to a harmless level.\\
Even if it would be possibile to get rid of U and Th contaminations, the \tnu\ decay induces an irriducibile background for the  \onu\ search.
The end-point of the \tnu\ spectrum (when neutrinos are emitted at rest)  contribute to the background in the ROI since all the available energy  is carried out by the two electrons expect for the two negligible  neutrino masses.  The ratio of \tnu\ to \onu\ event rate depends on the \tnu\ half life and on the energy resolution and could be assumed negligible in the case of IH region for an energy resolution better than one per cent  \cite{Artusa:2014wnl}.
On the other hand, accidental pile-up of two \tnu\ events  in the same detector  within a time window smaller than the typical time response of the detector can produce a signal that mimics a \onu\ decay.  This contribution can be suppressed  using the leading edge of the thermal response that range from microseconds (athermal sensors) to milliseconds (thermal sensors). 
 This turns out to be problematic only in the case of $^{100}$Mo which has the fastest observed \tnu\ decay and can contribute to ROI with a background of the level of   10$^{-2}$ \cpy
  \cite{Beeman:2011bg,Chernyak:2012zz,Chernyak:2014ska} in case of  slow NTD thermal sensors to 10$^{-4}$ \cpy in case of fast sensors sensitive to athermal phons as MMC \cite{Luqman:2016okt}.
The discrimination capability depends on the slope-to-noise ratio \cite{spieler} and it has been demonstrated that  the use of light detector with NLF signal amplification, could lower this  background  down to 6$\times$10$^{-5}$ \cpy \cite{Chernyak:2016aps}.

In the CUORE background budget \cite{Alduino:2017qet}, no positive indication of bulk contaminations in the other detector elements have been obtained, However, current upper limits could  translate to potentially dangerous counting rates for the CUPID background target. One order of magnitude improvement in the  sensitivities of presently screening technology is therefore mandatory.\\
The screening techniques commonly used are:  HPGe (High Purity Germanium detector), ICPMS (Inductively Coupled Plasma Mass Spectrometry, NAA (Neutron Activation Analysis). 
NAA and HPGe measurements can reach a sensitivity of the order of $\mu$Bq/kg  on $^{232}$Th in copper, the material used as supporting structure for the crystal absorber. The sensitivity is  limited by the mass of the copper sample that can't be increased {\it ad libitum} due to the self-absorption of the \ga\ lines inside the sample. It could be increased making use of pre-concentration of contaminants through chemical treatment of materials which is equivalent to a mass increase of the sample. The technique is  used in ICPMS measurements but can also be applied to  NAA or HPGe spectroscopy. However, it requires a dedicated study for each material as well as a very careful control of systematics.\\
Some detectors parts used in the form of foils, as super-insulation or flat cables, are not suitable for HPGe due to their small mass neither for NAA or ICPMS which have restrictive conditions on the material  that can be analyzed. In these cases, surface alpha spectroscopy through Si surface barrier diodes proved to reach competitive sensitivities. 
An alternative  technique consists of using a bolometric detector for the measurement of surface/bulk contamination; TeO$_2$ slabs can be used to realize a sandwich-like detector where samples are inserted in-between thin bolometers. Given the better energy resolution, the lower energy threshold and  the superior radio-purity \cite{Cardani:2012xq}, this approach  could reach a sensitivity up to a factor 100  higher than the actual ones and could provide information on the X-ray emission of the samples providing a complementary information for contamination identification.
\section{Other Rare Nuclear Decays and Processes}	
\label{rare}

\subsection{\al\ decays}	
The discovery of the \al\ decay of $^{209}$Bi with a half-life of 1.9$\times$10$^{19}$ yr in 2003 \cite{demarcillac} renewed the interest in the field of rare \al\ decays as a fundamental tool for the study of the structure of nuclei and  for a better understanding of the theoretical framework of nuclear models \cite{XuRen}.

The possibility to produce massive bolometers with a wide  choice of  materials has  very clear advantages. A significant amount of the nucleus of interest could be embedded in the detector itself, i.e.  the detector and the decay source coincide. As a consequence the decay is full contained in detector thus resulting in excellent detector efficiency. This aspect is of primary importance  for rare \al\ decays search also because of  the short (few microns) range  of a MeV \al\ particle in a solid medium.  In addition the high energy resolution and the capability to identify the interacting nature of the particle in the detector, as discussed  in section \ref{sb},  leads to tremendous background suppression, especially for rare  decays with  transition energy lower than 2.6 MeV (as for some lead isotopes) which would otherwise be overwhelmed by the near background from \ga\ emission of \TL.
Bolometers allow to measure half lives much longer than the age of the Universe. They led to the conclusive test on the identification  the \al\ decay of $^{209}$Bi \cite{demarcillac} and the discovery of its decays to the first excited level \cite{Beeman:2011kv,cardanirari}, to the discovery of $^{180}W$ ~\cite{Cozzini} and $^{151}$Eu ~\cite{Casali:2013zzr}  \al\ decays  and to set the most stringest lower limits on the half life of the \al\ decays of lead isotopes \cite{Beeman:2013Pb}.\\
An alternative approach consists in the  use of  a well known scintillating bolometer doped with the isotope under investigation. This allows to select a vey radio-pure crystal with high light yield and excellent \al/(\be/\ga) separation and to study more elements as  Sm, Nd, Os, Hf, Pt. The drawback is that the mass of the  candidate isotope is limited to few grams. This technique was recently proposed and used for a precise measurement of the half life and transition energy of the $^{148}$Sm \al\ decay  in a ZnWO$_4$ crystal \cite{Casali:2016vbw}.

\subsection{\be\ decays}	
The improvement of the sensitivity in the search for rare \al\ decays had an impact also on the study of extreme \be\ decay as the ones generated by small decay energies or  initial and final nuclear states with large angular momentum difference. \\
The knowledge of rare \be\ decay existence and their spectral shape is  fundamental since they can represent a background for the \onu\ search. The $^{214}$Bi, for example,  has a $Q_\beta$= 3270 keV thus exceeding the \QBB\ of  the more studied \onu\ isotopes (see section \ref{err}). In about 19\% of the cases it decays to the ground state of $^{214}$Po with change in angular momentum J and parity $\pi$, $\Delta J^{\Delta \pi}=1^-$ (first forbidden non-unique transition). Its shape is not well measured experimentally neither  theoretically predicted, moreover it must be  taken into account that  forbidden \be\ shape  can significantly deviate from  known allowed spectra.
 
 But there is a more important feature related to forbidden \be\ decays: their shape  could be used to infer  the ratio of the  weak axial to vector coupling constant g$_A$/g$_V$ in nuclear decays.\\
 The \onu\ half life is proportional to the fourth power of  g$_A$. The recent analyses of nuclear models in \be\ and \tnu\ decays indicate  that the value of g$_A$ could be quenched, up to a ratio of g$_{free}$/g$_A$ $\sim$4, where g$_{free}$ = 1.27 is the free value of g$_A$ inferred from the neutron decay. This could potentially translates into a two order of magnitude difference in the sensitivity of \onu\ experiment. This {\it naive} expectation has been very recently  scaled back to a factor between two and six if a consistent approach is used for the calculation of the \tnu\ and \onu\ decays \cite{Suhonen:2017rjf}. 
 Nevertheless the measurement of g$_A$ is of pivotal importance. This could be inferred by the shape of $\Delta J^{\Delta \pi}=4^+$ non-unique forbidden \be\ decays as for $^{113}$Cd and $^{115}$In \cite{Haaranen:2016rzs,Haaranen:2017ovc,Kostensalo:2017xxq,Kostensalo:2017jgw,Suhonen:2017krv}. For such decays  the shape of energy spectrum relies on the sum of different nuclear matrix elements with different phase space factors which include $g_A$ and $g_V$ and their values could be extracted by the comparison between  theoretical and  experimental spectra. 
 While in the case $^{113}$Cd the spectral shape has been characterized  \cite{Belli}, only and old measurement \cite{Pfe} exists for the shape of $^{115}$In. To perform a clean and reliable measurement a  10 g LiInSe$_2$ scintillating bolometer  is currently  in data taking at the Modane underground laboratory \cite{Tretyak:2017zqd}.
 
\subsection{Electron capture processes}	
Bolometers can play in important role also in the search of other rare nuclear processes as the rare electron capture. When the source of the decay is embedded in the bolometer, in fact,  a signal corresponding to the total binding energy of the captured electron can be measured with very high efficiency because X-rays/Auger electrons following the atomic de-excitation are fully contained. Moreover the excellent energy resolution is a powerful  tool to discern externally generated \ga\ from  X-ray and  electron cascades. \\
As example the electron capture of $^{123}$Te is predicted but not yet observed. The best limit, obtained with a TeO$_2$ bolometer is $T_{1/2}>5.0\times10^{19}$ y \cite{Alessandrello:2002ag}. A previous observation \cite{Alessandrello:1996zz} was confuted and explained  as  the electron capture in $^{121}$Te, an isotope created by neutron capture on the 0.09\% natural abundant $^{120}$Te isotope at see level.
The importance of this measurement relies on the fact that it could be used to constrain and test nuclear models used to estimate intensities  for rare electroweak decays \cite{civitarese}, models that in some case foresee a suppression of the rate up to six orders of magnitude \cite{broglia97}.  
The CUORE experiment, with is huge mass compared to its predecessors, will be able to improve the results by orders of magnitude and possibly to discover the electron capture of $^{123}$Te.

\section{Conclusions}
Bolometers are cryogenic calorimeters which base their principle on the phonon detection. They exhibit: excellent energy resolution, low energy threshold, high detection efficiency, wide choice of material for the  calorimeter absorber. These characteristics make them 
 one of the best performing instruments in several fields: double beta decay search, neutrino mass measurement, dark matter search, CMB precision measurement, high resolution X-ray detection, rare nuclear process detection. \\
 This review is focused on the bolometric applications in the field of rare nuclear processes: in particular on the neutrinoless double beta decay search.
The demand for the  increase of the experimental sensitivity  imposes a series of technical challenges and improvements of the actual technology. In particular, experiments  aiming at covering the inverted hierarchy region of the neutrino mass scheme and possibly  discovering the Majorana neutrino nature, need to lower the background in the region of interest  to the level of 10$^{-4}$ \cpyI and increase the source mass. This implies a manifold effort: the development of passive methods for background reduction and new screening techniques, the growth of very radio-pure enriched crystals and the implementation of reliable active background rejection techniques. \\
The first pilot demonstrators, using enriched scintillator bolometers for particle identification, are already in data taking while the development of detectors for the Cherenkov light detection in TeO$_2$ bolometers is rapidly growing and entering into its final phase.
On the other hand, the successful operation of the CUORE experiment (988 massive bolometers) ensures that hundred of kgs of isotopes can be studied in a stable and reliable cryogenic system.\\
Despite three decades have passed  since they have been conceived, bolometers are still a very active field and performances are continuously improving. The viability of a next generation experiment in an almost  background-free environment, is within the reach if actual R\&D will be successful.\\
Moreover, the superior bolometric features renewed in the last five years the interest in rare nuclear processes as a tool for the comprehension of nuclear models.  The improvements in terms of performances and radio-purity of material, requested by  the neutrinoless double beta decay search, are beneficial for all the rare nuclear process searches and will boost the sensitivity to unprecedented levels. \\

\bibliographystyle{ws-ijmpa}
\bibliography{bellini}

\end{document}